\pdfoutput=1
\documentclass[conference]{IEEEtran}

\usepackage[cmex10]{amsmath}
\usepackage{amsthm}
\usepackage{amssymb}
\usepackage{mathrsfs}
\usepackage{graphicx}
\usepackage{float}
\usepackage{array}
\usepackage{epstopdf}
\usepackage{multirow}
\usepackage{cite}
\usepackage{color}

\begin{document}

\title{COVID CT-Net: Predicting Covid-19 From Chest CT Images Using Attentional Convolutional Network}

\author{Shakib Yazdani$^1$,  Shervin Minaee$^2$, Rahele Kafieh$^3$, Narges Saeedizadeh$^3$, Milan Sonka$^4$\\
$^1$ECE Department, Isfahan University of Technology, Iran\\
$^2$Snap Inc., Seattle, WA, USA\\ 
$^3$Medical Image and Signal Processing Research Center, Isfahan University of Medical Sciences, Iran \\ 
$^4$Iowa Institute for Biomedical Imaging, The University of Iowa, Iowa City, USA \\ 
}

\maketitle

\begin{abstract}
The novel corona-virus disease (COVID-19) pandemic has caused a major outbreak in more than 200 countries around the world, leading to a severe impact on the health and life of many people globally. As of Aug 25th of 2020, more than 20 million people are infected, and more than 800,000 death are reported.
Computed Tomography (CT) images can be used as a as an alternative  to  the  time-consuming ``reverse transcription polymerase chain reaction (RT-PCR)'' test, to detect COVID-19. 
In this work we developed a deep learning framework to predict COVID-19 from CT images.
We propose to use an attentional convolution network, which can focus on the infected areas of chest, enabling it to perform a more accurate prediction. 
We trained our model on a dataset of more than 2000 CT images, and report its performance in terms of various popular metrics, such as sensitivity, specificity, area under the curve, and also precision-recall curve, and achieve very promising results.
We also provide a visualization of the attention maps of the model for several test images, and show that our model is attending to the infected regions as intended.
In addition to developing a machine learning modeling framework, we also provide the manual annotation of the potentionally infected regions of chest, with the help of a board-certified radiologist, and make that publicly available for other researchers.
\end{abstract}

\IEEEpeerreviewmaketitle

\section{Introduction}
The coronavirus disease 2019 (COVID-19), caused by severe acute respiratory syndrome coronavirus 2 (SARS-CoV-2), is an ongoing pandemic. By August 27, more than 24 million confirmed cases, and more than 830,000 deaths cases were reported in the world \cite{Radio1}. This has led to great public health concerns, and World Health Organization (WHO) declared the outbreak to be a Public Health Emergency of International Concern (PHEIC) on January 30, 2020 and recognized it as a pandemic on March 11, 2020 \cite{Radio1-1, Radio1-2}.

For diagnosis of COVID-19, real-time reverse transcription polymerase chain reaction (RT-PCR) is regarded as the reference standard; however, recent studies addressed the importance of chest computed tomography
(CT) examination in COVID-19 patients with false negative
RT-PCR results \cite{Radio1-3}.
In initial stages of COVID-19, bilateral multilobar ground-glass opacification (GGO) are common with a peripheral or posterior distribution, mainly in the lower lobes. In intermediate stage of disease, an increase in the number and size of GGOs and progressive transformation of GGO into multifocal consolidative opacities are visible day 10 after the symptom onset \cite{Radio1-4}. 
Imaging patterns corresponding to clinical improvement of COVID-19 usually occur after week 2 of the disease and include gradual resolution of consolidative opacities and decrease in the number of lesions and involved lobes \cite{Radio1-4}.
Chest CT shows a low rate of missed diagnosis of COVID-19 (3.9\%, 2/51) and may be useful as a standard method for the rapid diagnosis of COVID-19 \cite{Radio1-5}.

\begin{figure}[h]
\begin{center}
  \centerline{\includegraphics[width=0.96\linewidth]{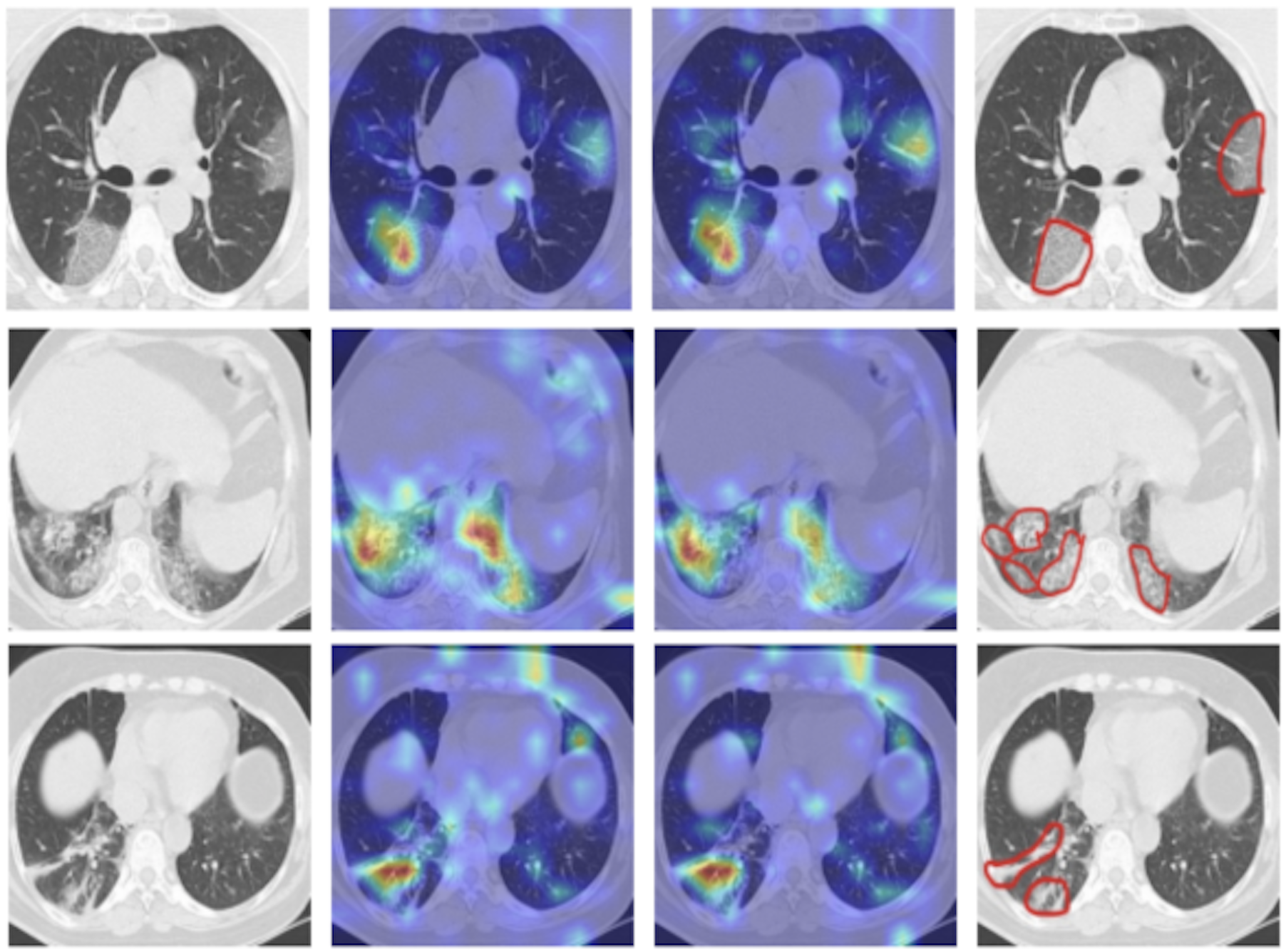}}
\end{center}
   \caption{The attention map of the model for three sample COVID-19 images. We also provide the manually detected infected regions by a board-certified radiologist. As we can see the attention maps are mostly active around the infected regions.}
\label{intro_attentionmap}
\end{figure}

Artificial Intelligence can help COVID-19 community by accurate segmentation of infections and automatic diagnosis and classification of X-ray and CT images. A number of studies aimed to separate COVID-19 patients from non-COVID-19 subjects (that include common pneumonia subjects and non-pneumonia subjects) \cite{Radio1-6}.
Chen et al. \cite{Radio1-7} used CT images from 51 COVID-19 patients and 55 patients with other diseases to train a UNet++ based segmentation model for segmenting COVID-19 related lesions. They finally labeled each image based on the segmented lesions and reached 95.2\% (Accuracy), 100\% (Sensitivity), and 93.6\% (Specificity). 
Jin et al. \cite{Radio1-8} used chest CT images from 496 COVID-19 patients with 1385 negative cases. They first segmented the lung using a 2D CNN based model and then found slices related to COVID-19. The model achieved sensitivity of 94.1\%, specificity of 95.5\%, and AUC of 0.979.
In \cite{Radio1-9}, a similar strategy was taken in 3D format using U-Net+3D CNN based model (DeCoVNet) for segmentation and 3D CNN for classification. The model achieved a sensitivity of 90.7\%, specificity of 91.1\%, and AUC of 0.959.
DeepPneumonia \cite{Radio1-10} was proposed using ResNet50 to detect patients with COVID-19 (88 patients) from bacteria pneumonia patients (101 patients) and 86 healthy people. They reached an accuracy of 86.0\% for pneumonia classification (COVID-19 or bacterial pneumonia), and an accuracy of 94.0\% for pneumonia diagnosis (COVID-19 or healthy).

In this work, we propose a deep learning framework to predict COVID-19 from the chest CT images directly. 
We propose to use an attentional residual convolutional network, which can better focus on the infected areas, leading to an accurate classification of COVID-19 cases.
We train our model on a public dataset of COVID-19 CT images. Through experimental results, we show that the attention maps of this model are mostly active around the infected regions, which is really encouraging. 
This is shown for three sample COVID-19 images in Figure \ref{intro_attentionmap}.

We also perform a thorough experimental study, providing both quantitative and qualitative analysis.
In terms of quantitative study, we report the model's sensitivity, specificity, receiver operating characteristic, precision-recall curve, as well as the distribution of predicted scores for both healthy and COVID-19 images. 
In terms of qualitative results, we visualize the attention maps of the trained model on some of the COVID-19 test image, and also provide the potentially infected regions by a visualization technique based on sliding window on the model predictions, and show that our model is mostly sensitive to the infected regions.

Here are the contribution of this paper:
\begin{itemize}
    \item We proposed a framework based on attentional convolutional network, to predict COVID-19 from CT images.
    \item We provided a detailed analysis of the model, by looking at various qualitative and quantitative metrics, as well as comparison with baselines.
    \item We provide the visualization of the attention maps in the trained model, and show that the model is mostly sensitive to the infected areas.
    \item We also annotate the infected regions for a subset of COVID-19 images in this dataset with the help of a board-certified radiologist, and make them available publicly for other researchers to use them.
\end{itemize}

The structure of the rest of this paper is as follows.
Section \ref{sec_method} provides the details of the proposed framework. 
In \ref{sec_dataset}, we provide an overview of the dataset used in this work, and its characteristics.
Section \ref{sec_result} provides the experimental studies and comparison with previous works.
And finally the paper is concluded in Section \ref{sec_conc}.

\section{The Proposed Framework}
\label{sec_method}
Deep learning based models have achieved promising results in various computer vision and medical image analysis problems in the past few years and have shown to beat classical statistical models based on hand-crafted features with a large margin in several studies \cite{med1, med2, med3, med4, med5, med6}.
They have also been used by several research groups recently, to detect and segment COVID-19 from medical images \cite{cov1, cov2, cov3, cov4}.
In this work we focused on a predictive model based on attentional covnolutional networks, to detect COVID-19 from CT images.
Given the fact that COVID-19 infects local regions of Lung, we used a CNN model with attention mechanism to better focus on the infected areas, resulting in higher accuracy.
Our model architecture is similar to residual attentional convolutional network, and its details is provided in the next part.

\subsection{Residual Attention Network}
\begin{figure*}[h]
\begin{center}
   \includegraphics[page=2,width=0.99\linewidth]{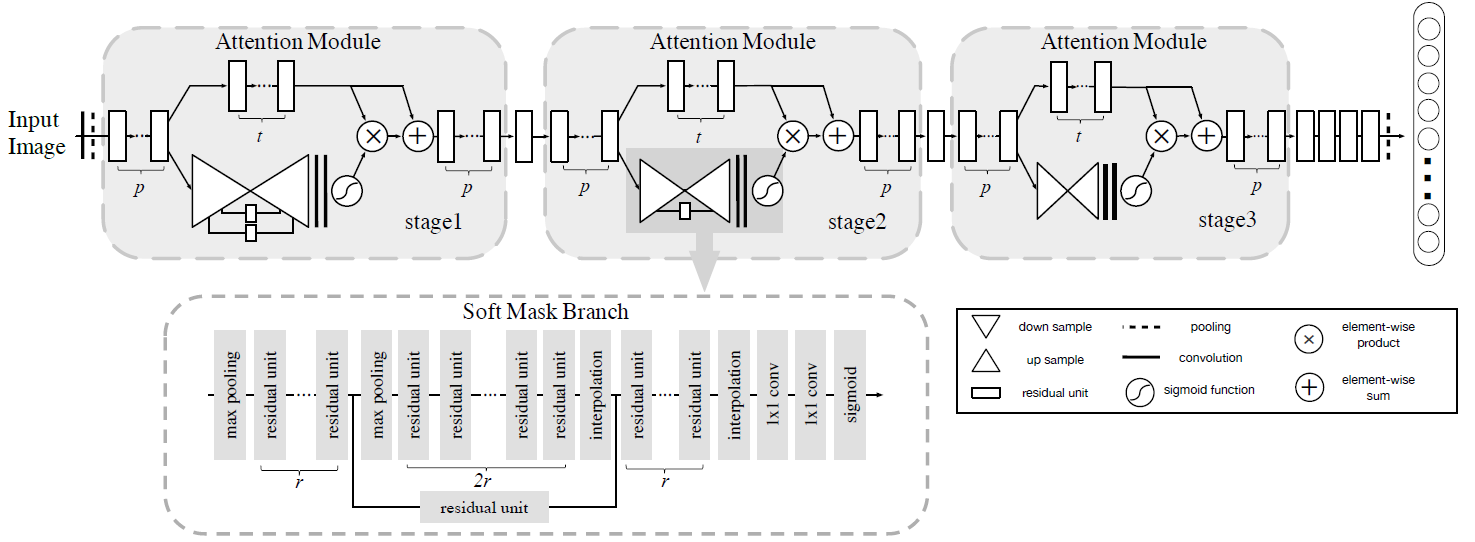}
\end{center}
   \caption{Illustration of Residual Attention Network. In this network three hyperparameters are used which are p, t and r. The hyperparameter \emph{p} represents the total number of pre-processing residual units used before splitting into trunk and mask branches. \emph{t} and \emph{r} denote the number of residual units in trunk branch and the number of neighboring pooling layers in the mask branch respectively. We use the following hyper-parameters setting: \{p = 1, t = 2, r = 1\} in our experiments. Courtesy of \cite{res-attention}. }
\label{res_attention}
\end{figure*}

We adopt a residual attention model as our model architecture, which is a combination of multiple stacked residual modules empowered with attention-aware units. 
Each attention module is composed of two branches: trunk branch and mask branch. The trunk branch which is the upper branch of attention module is supposed to do the feature extraction and it can be based on a single block of any convolutional neural network  \cite{res-attention}.
The  architecture of our model is shown in Figure \ref{res_attention}. 

In this work, we use a single block of ResNet-56 as our attention module basic unit. Considering input \emph{x} the output of trunk branch \emph{T(x)} can be weighed using mask branch with the output of \emph{M(x)}, which is based on a encoder-decoder architecture. 
Therefore, the output of each attention module can be obtained as in Eq \ref{attention_module_H1}:
\begin{equation}
\begin{aligned}
&  H_{i,c}(x) =  M_{i,c}(x) * T_{i,c}(x)
\end{aligned}
\label{attention_module_H1}
\end{equation}
where i ranges over all spatial positions and c $\in$  \{1, . . . , C\} 
is the index of the channel. 
 
However, naive stacking of attention modules can lead to performance drop because dot production with mask range from zero to one will decrease the value of features in deep layers. The ability to train a complex neural network such as the aforementioned model that uses an stacking architecture is the intuition behind using a residual attention learning. Thus, the modified attention module output \emph{H} can be written as follows:
\begin{equation}
\begin{aligned}
&  H_{i,c}(x) =  (1 + M_{i,c}(x)) * F_{i,c}(x)
\end{aligned}
\label{attention_module_H}
\end{equation}
where \emph{M(x)} ranges from [0,1], and if \emph{M(x)} is set to 0 we can obtain the original features \emph{F(x)}. 

Additionally, the difference between the original ResNet and ours lies on the way residual learning is used. In the former one, residual learning is written as $H_{i,c}(x)$ =  $x$ + $F_{i,c}(x)$, where $F_{i,c}(x)$ estimates the residual function, and in the latter one, $F_{i,c}(x)$ is the features generated by our baseline CNN.

\begin{figure}[h!]
\begin{center}
   \includegraphics[page=2,width=0.99\linewidth]{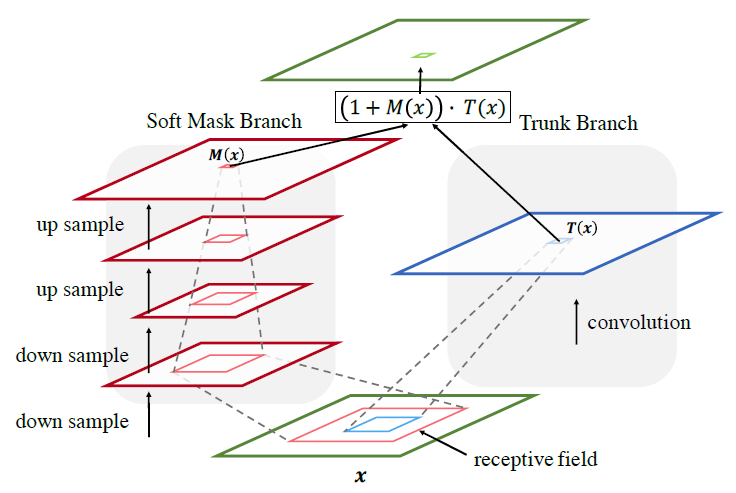}
\end{center}
   \caption{The identical mapping between the trunk branch and the mask branch. Courtesy of \cite{res-attention}.}
\label{receptive_field}
\end{figure}

Soft mask branch is consisted of a bottom-up top-down fully convolutional network. In the bottom-up approach, it first extracts the global features of the whole image, and then it combines the extracted global features with the original feature maps using the top-down architecture.
First, max pooling is performed to down sample the input image, following that several number of residual units are used. For the top-down part, the same number of linear interpolation (as max pooling) is used to not only keep the output size the same as input feature map but also to downsample them. Finally, following two consecutive 1x1 convolution layers, a Sigmoid activation function is applied to normalize the output between $[0,1]$.
A more understandable diagram of the identical mapping between the trunk branch and the mask branch is shown in Figure \ref{receptive_field}.

\section{COVID-19 CT Dataset}
\label{sec_dataset}
This study is performed on the publicly available SARS-CoV-2 CT scan dataset, which has been collected from real patients in the hospitals from Sao Paulo, Brazil \cite{dataset}.
It contains 2482 CT scan images in total, out of wchich 1252 are labeled as COVID-19, and 1230 belong to the people not infected by SARS-CoV-2. 
This data is divided into the training and test sets, with the exact number of samples provided in Table \ref{tab:dataset_split}.

As the number of samples is important for a model to be trained well, we used data augmentation technique to increase the number of training samples (via Augmentor library in Python \cite{Aug}).
Different transformations are used for data augmentation, such as  rotations, flip left-right, flip top-bottom, random-distortion, and skew-titl.
We were able to increase the number of training samples by a factor of 4, with the help of data augmentation. 

\begin{table}[ht]
\centering
  \caption{Train/Test of the  dataset with and without Data Augmentation}
  \centering
\begin{tabular}{|c|c|c|c|c|}
\hline
Data& \multicolumn{2}{c|}{No Augmentation} &  \multicolumn{2}{c|}{with Augmentation}\\\hline
&COVID-19&non-COVID-19&COVID-19&non-COVID-19\\\hline
Training  & 764 & 718 & 3056 & 2868 \\\hline
Test  & 488& 512& 488& 512\\ \hline
Total  & 1252 &1230&3544 &3380 \\ \hline
\end{tabular}
\label{tab:dataset_split}
\end{table}

\begin{figure}[h!]
\begin{center}
   \includegraphics[page=2,width=0.99\linewidth]{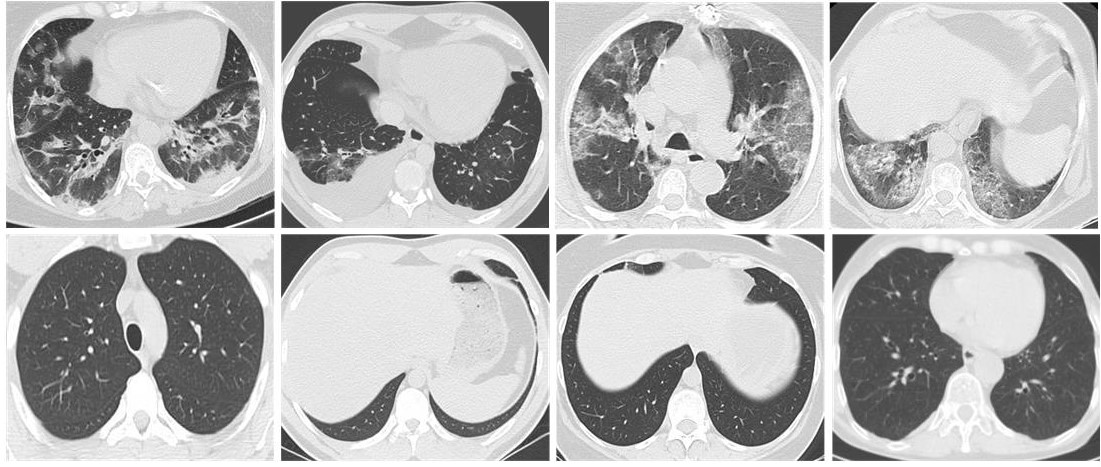}
\end{center}
   \caption{Eight sample images from this dataset. The images in the first row belong to the covid-19 cases, and the images in the second row belong to the non-covid cases.}
\label{data_samples}
\end{figure}

Eight sampels images from this dataset are shown in Figure \ref{data_samples}, from which four of them belong to COVID-19 patients, and the other four to the none-COVID cases.

\section{Experimental Results}
\label{sec_result}
In this section we provide a detailed analysis of experimental studies performed for evaluating our classification framework.
We present both qualitative and quantitative results, and also compare our results with some baseline.
Before getting into the result details, we will provide a brief overview of popular metrics used for evaluating classification models, and also discuss about the hyper-parameters choice in our study.

\subsection{Evaluation Metrics }
We used two widely adopted metrics, for evaluating the performance of medical image classification models, sensitivity and specificity, as defined below:

\begin{equation} \label{sensitivity}
\begin{aligned}
\text{Sensitivity (Sen.)} & = \frac{TP}{TP+FN} \;\; 
\end{aligned}
\end{equation}
\begin{equation} \label{Specificity}
\begin{aligned}
\text{Specificity(Spec.)} & = \frac{TN}{TN+FP} \;\; 
\end{aligned}
\end{equation}
where, TP (True Positive) is the number of correctly classified images of a class, FP (False Positive) is the number of the wrong classified images of a class, FN (False Negative) is the number of images of a class that have been detected as another class, and TN (True Negative) is the number of images that do not belong to a class and were not classified as belonging to that class.
In a more medical-friendly language, test sensitivity is the ability of a test to correctly identify those with the disease (true positive rate), whereas test specificity is the ability of the test to correctly identify those without the disease (true negative rate).

Besides these metrics, we also look at the model performance in terms of ROC-curve, precision-recall curve, an histogram of predicted scores.
The receiver operating characteristic curve (ROC curve ) shows the trade-off between the true positive rate (TPR) and the false positive rate (FPR) at various threshold settings. The true-positive rate is also known as sensitivity, recall, or probability of detection in machine learning. The false-positive rate is also known as the probability of false alarm and can be calculated as (1 − specificity). The area under the ROC curve is AUC.
The precision-recall curve is created by plotting precision against Recall for different threshold. A high area under the curve represents both high recall and high precision, where high precision relates to a low false positive rate, and high recall relates to a low false negative rate. High scores for both show that the classifier is returning accurate results (high precision), as well as returning a majority of all positive results (high recall).

\subsection{Model Hyper-parameters Impact}
Hyper-parameters play an important role in the performance of machine learning models, specially deep neural networks, and need to be tuned carefully. 
Hyper-parameter tuning can be done both automatically and manually. In our work, we manually tried several combinations of different hyper-parameters so that we can select the best option. 
We fixed the number of epochs to 100 and batch size to 16, and varied the other hyper-parameters. 
Form optimizer, RMSProp has shown to give us the best result, and for learning rate we achieved the best validation accuracy with a rate of 0.01. It is worth mentioning that although momentum-based nesterov accelerated stochastic gradient descent reaches an acceptable accuracy of 0.907, takes a long time to converge in comparison to other below optimisers.
Our entire implementation is done in Keras with Tensorflow backend.

Table \ref{tab:Different Optimizer} shows the impact of different learning rates, and different optimizers on the model validation accuracy.
 
 \begin{table}[h!]
\centering
  \caption{The model  performance for different optimizers, and different learning rates.
  }
\centering
 \begin{tabular}{ |p{1.10cm}|p{1cm}|p{1cm}|p{1.75cm}|}
 \hline
Loss&	Optimizer&	Learning  
Rate&	Accuracy \ \
\\
 \hline
&   ADAM  & 0.001	 &	0.907 $\pm$ 0.001\\ \cline{2-4}
&	SGD	& 0.001 &	0.885 $\pm$ 0.002 \\ \cline{2-4}
 BCE & 	& 0.01&	0.920 $\pm$ 0.001\\ 
&RMSProp	& 0.001&	0.881 $\pm$ 0.002\\
&	& 0.0001&	0.900 $\pm$ 0.001	\\

 \hline
 \end{tabular}
 \label{tab:Different Optimizer}
\end{table}

\subsection{Predicted Scores}
\begin{figure}[h!]
\begin{center}
   \includegraphics[page=2,width=0.9\linewidth]{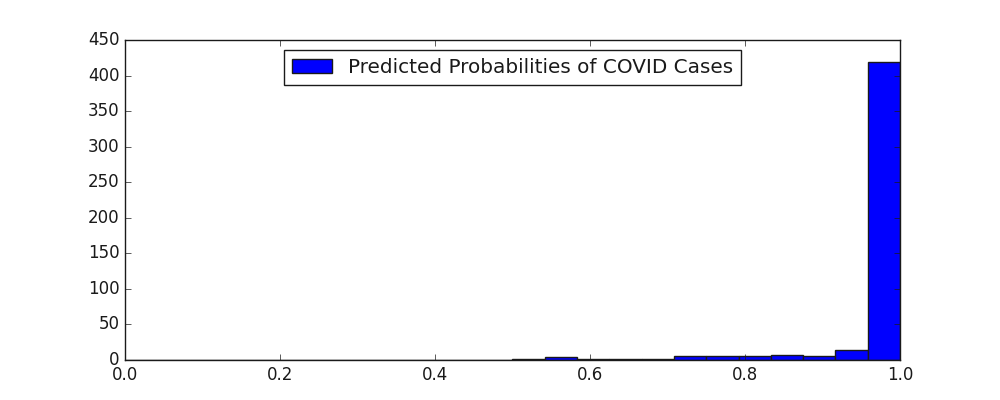}
\end{center}
   \caption{The predicted probability scores of COVID cases on the test set}
\label{Pred_Prob_of_COVID_Cases}
\end{figure}

\begin{figure}[h!]
\begin{center}
   \includegraphics[page=2,width=0.9\linewidth]{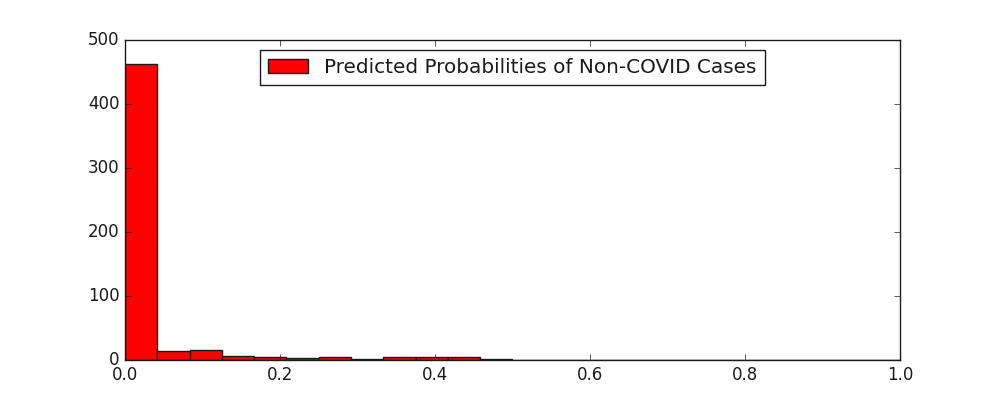}
\end{center}
   \caption{The predicted probability scores of Non-COVID cases on the test set}
\label{Pred_Prob_of_Non-COVID_Cases}
\end{figure}

As mentioned earlier, our model predicts a probability score which shows the likelihood of a sample belonging to COVID-19 class. The higher this score, the more likely the image to have COVID-19 infection.
In this section we present the distribution of the predicted scores on the test set. 
Ideally we expect the model to predict a very high likelihood score (close to 1) for COVID-19 patients, while predicting low scores (close to 0) for healthy people. 
Figures \ref{Pred_Prob_of_COVID_Cases} and \ref{Pred_Prob_of_Non-COVID_Cases} show the histogram of the predicted probabilities of both COVID-19 and Non-COVID cases on the test set.
As it can be seen from these figures, our model does a good job of separating the scores of samples from each class.

\subsection{Cut-off Threshold Impact on Model Performance}
As it is clear, in order to decide if a sample is COVID-19, we can compare its predicted score with a cut-off threshold and if it is larger than threshold it will be labeled as COVID-19. 
Therefore, different cut-off thresholds can result on a different set of samples being considered as COVID-19, and this can lead to a trade-off between false positives and false negatives rates. The default cut-off threshold is usually set to 0.5, but that does not guarantee to achieve the best results. 
In order to find the best threshold value, we try different thresholds ranging from 0.1 to 0.9, and provide their performance in table \ref{residual-attention-thresh}.

\begin{table}[h!]
\centering
  \caption{Sensitivity, specificity scores of the proposed model for different threshold values. The confidence scores are also provided, to see the statistical significance of relative gains.
  }
  \centering
\begin{tabular}{|c|c|c|c|}
\hline
Threshold  & Sensitivity & Specificity & F1 score\\
\hline
0.1&	0.850 $\pm$ 0.002&	0.962 $\pm$ 0.001 & 0.900 $\pm$ 0.001\\ \hline
0.2&	0.870 $\pm$ 0.002&	0.951 $\pm$ 0.001 & 0.906 $\pm$ 0.001\\ \hline
0.3&	0.893 $\pm$ 0.002&	0.945 $\pm$ 0.001 & 0.915 $\pm$ 0.001\\ \hline
0.4&	0.893 $\pm$ 0.001&	0.939 $\pm$ 0.001 & 0.913 $\pm$ 0.001\\ \hline
0.5&	0.903 $\pm$ 0.001&	0.935 $\pm$ 0.001 & 0.916 $\pm$ 0.001\\ \hline
0.6&	0.893 $\pm$ 0.001&	0.939 $\pm$ 0.001 & 0.913 $\pm$ 0.001\\ \hline
0.7&	0.893 $\pm$ 0.001&	0.945 $\pm$ 0.001 & 0.915 $\pm$ 0.001\\ \hline
0.8&	0.870 $\pm$ 0.002&	0.951 $\pm$ 0.001 & 0.906 $\pm$ 0.001\\ \hline
0.9&	0.850 $\pm$ 0.002&	0.962 $\pm$ 0.001 & 0.900 $\pm$ 0.001\\ \hline
\end{tabular}
\label{residual-attention-thresh}
\end{table}

\subsection{Precision-Recall Curve, ROC Curve, and Confusion Matrix}
\begin{figure}[h!]
\begin{center}
   \includegraphics[page=2,width=0.9\linewidth]{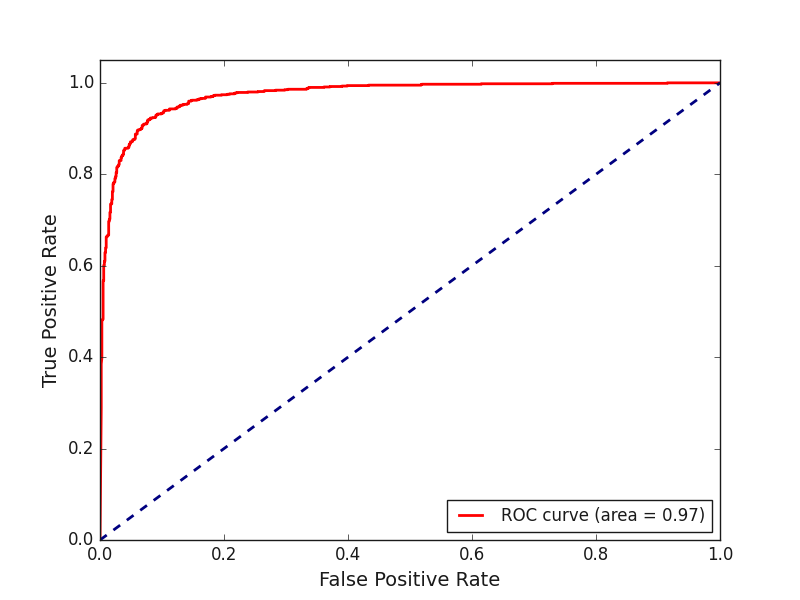}
\end{center}
   \caption{Receiver Operating Characteristic curve for the test set}
\label{roc-curve}
\end{figure}

Evaluating a model only by sensitivity and specificity metrics is difficult because these metrics are highly dependent on the threshold value that is used to tune the model performance. 
When we are dealing with a binary classification problem, we are concerned with two types of errors, false positive rate, and false negative rate. 
A better way to evaluate a model is to use some metrics which combines FPR and FNR, and gives us  a global picture of the model. 
A common way to do so is the Receiver operating characteristic (ROC), which shows the true positive rate (y-axis) as a function of the false positive rate (x-axis) for various thresholds ranging from zero to one. 
The ROC curve of our model can be seen in Figure \ref{roc-curve}. 

\begin{figure}[h!]
\begin{center}
   \includegraphics[page=2,width=0.9\linewidth]{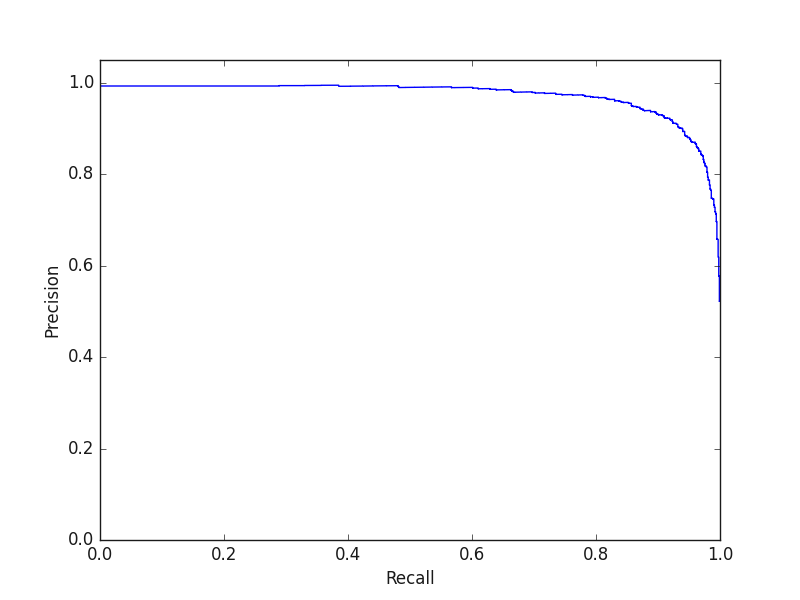}
\end{center}
   \caption{Precision-Recall curve for the test set}
\label{precision-recall-curve}
\end{figure}

Another way to compare the performance of a model for all possible thresholds is through the precision-recall curve. Precision is defined as a ratio of the number of true positives divided by the sum of the positives. 
Recall or sensitivity is also calculated as a ratio of the number of true positive images divided by the sum of the total number of images regarded as true positive and false negative. 
The precision-recall curve of our model is illustrated in Figure \ref{precision-recall-curve}.

\begin{figure}[t]
\begin{center}
   \includegraphics[page=2,width=0.7\linewidth]{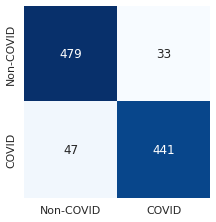}
\end{center}
   \caption{Confusion Matrix Plot}
\label{cm-matrix}
\end{figure}

Furthermore, the exact number of images labeled either as COVID-19 cases or Non-COVID-19 cases in the test set can be seen in the confusion matrix provided in Figure \ref{cm-matrix}.

\subsection{Visualization of Attention Layers}
To visualize the attention maps, we make use of Grad-CAM and also its generalized version called Grad-CAM++. 
Both of these methods are used as a visualization technique to calculate which of the spatial locations in convolutional layers are more important than others. 
This is done through computing the gradient of the output class prediction with respect to the feature maps of the desired convolutional layer. Grad-CAM++ makes use of a different derivation to address the shortcoming of the popularly used Grad-CAM, such as multiple occurrences of a class in an image  \cite{grad_cam}, \cite{grad_cam+}.

\begin{figure}[h!]
\begin{center}
  \centerline{\includegraphics[width=0.99\linewidth]{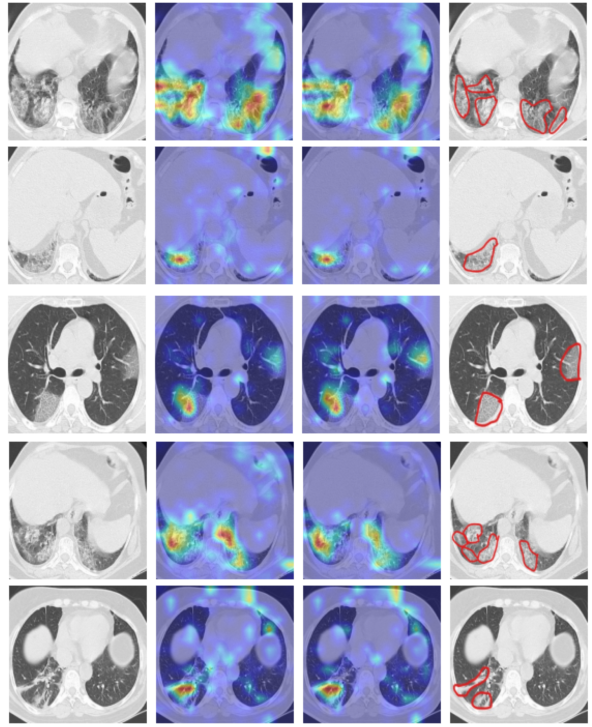}}
\end{center}
   \caption{Visualization of the third attention block. The left-most column represents the original image, the second and third columns denote the attention maps using Grad-CAM method, and  Grad-CAM++ respectively, and the last column shows the the annotated image by our radiologist.}
\label{third-attention-block}
\end{figure}

The visual generated by Grad-CAM and Grad-CAM++ of the third attention block layer of our model, for five sample images are shown in Figure \ref{third-attention-block}.
With the help of a board-certified radiologist, we also provide the manual regions which are likely to be infected with COVID-19 in these images two, and show them here. 
As we can see our model is very active around the infected areas, which is really encouraging.

\subsection{Heatmap of Infected Regions by Model}
\begin{figure}[h!]
\begin{center}
  \centerline{\includegraphics[width=0.99\linewidth]{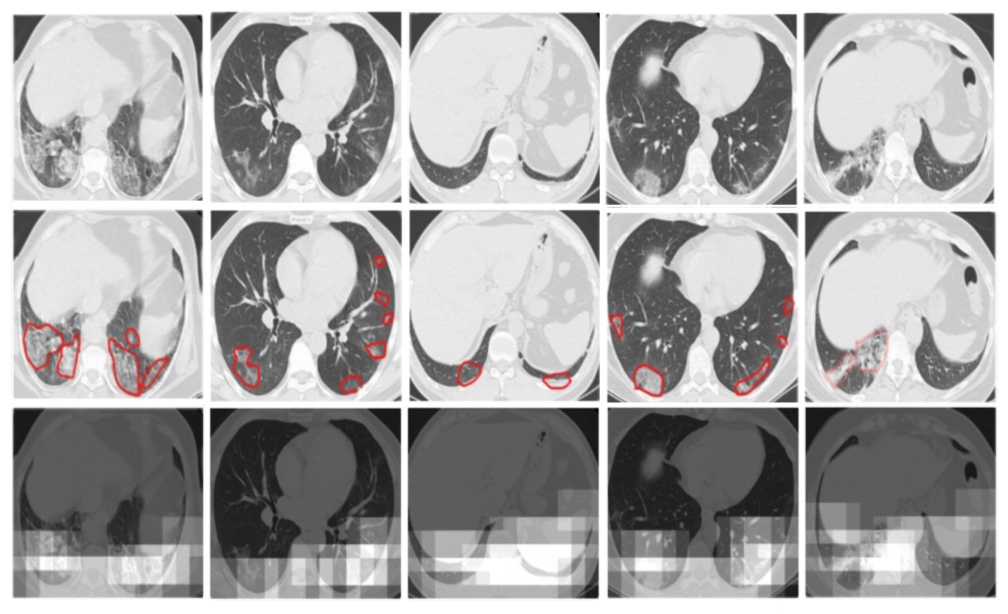}}
\end{center}
   \caption{The heatmap of the potentially infected regions, starting from the left, the first image is the original in the dataset and following that the heatmap of infected regions can be seen and the top most right one is the annotated image by our radiologist.}
\label{potentially-infected-reg}
\end{figure}

We also used another simple technique (inspired by the work of Zeiler and Fergus \cite{fergus}) to show the potentially infected regions of each image, with a sliding window technique on the model predictions. 
This approach works by zeroing out an NxN patch in the original image, finding its impact on the model's prediction and seeing if that lead the model to make a mistake. If by occluding that region our model incorrectly classifies a COVID-19 image as Non-COVID, that region should be regarded as a potentially infected region in our CT images. On the other hand, if this is not the case and the model performance is not affected by occluding a region, then we treat that region as not infected. 
Once we repeat this process for various sliding windows of size NxN (by shifting them with a stride of S each time), we can plot a saliency map of the potentially infected regions for COVID-19. 
Here we provide the detected regions for 5 sample images from our test set in Figure \ref{potentially-infected-reg}.
As we can see the model mostly contains the manually detected infected regions.

\subsection{Training Convergence Analysis}
\begin{figure}[h!]
\begin{center}
   \includegraphics[page=3,width=0.9\linewidth]{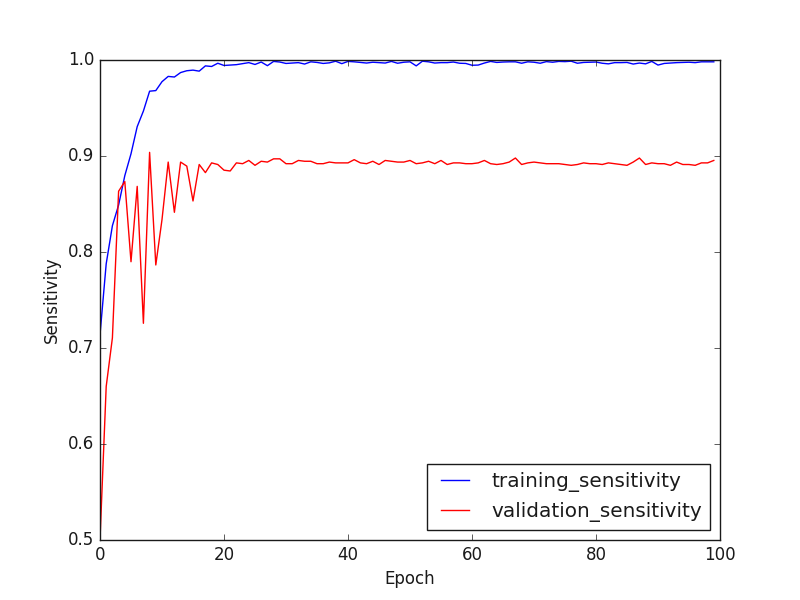}
\end{center}
   \caption{The training and validation sensitivity of our model}
\label{sensitivity_val}
\end{figure}

\begin{figure}[h!]
\begin{center}
   \includegraphics[page=3,width=0.9\linewidth]{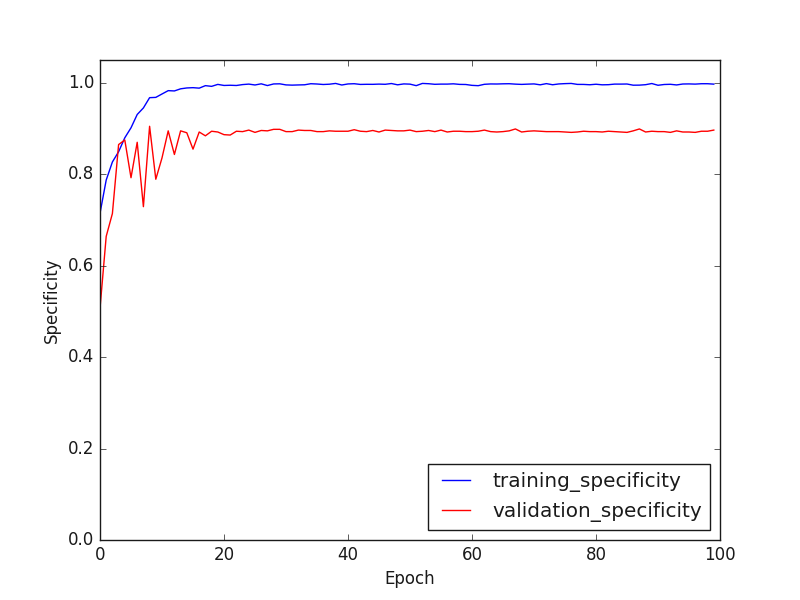}
\end{center}
   \caption{The training and validation specificity of our model}
\label{specificity}
\end{figure}

To see the model convergence during training, we provide the recall (sensitivity), and  specificity of the model on different epochs in Figure \ref{sensitivity_val} and Figure \ref{specificity} respectively. 
Its worth mentioning that, the default threshold value of 0.5 is used in these figures.

\section{Conclusion}
\label{sec_conc}
In this work, we proposed a deep learning framework for COVID-19 prediction from chest CT images. 
To enable our model to better attend to the infected parts of the images, we used attention convolutional network. 
We train our model one of the largest publicly available (so far) covid-19 CT image datasets, and provide a detailed experimental study, by looking on model performance in terms of sensitivity, specificity, precision-recall curve, and ROC curve.
In addition to the quantitative analysis, we also provide a visualization of the model attention maps and show how similar they are with the manually detected infected regions by our board-certified radiologist.
As a part of this work, we also make the manually detected infected regions publicly available, for the future research use by the community. 

\section*{Acknowledgment}
The authors  would like to thank the providers of the SARS-CoV-2 CT scan dataset, for making it publicly available.
We would also like to thank our radiologist, Doctor Ghazaleh Soufi, for her her help in manually detecting the infected regions of CT images, which were not available as a part of the original dataset.

\end{document}